\documentclass[a4paper]{jpconf}

\usepackage{graphicx}
\usepackage{dcolumn}
\usepackage{bm}

\usepackage{amsmath}
\usepackage[latin1]{inputenc}

\newcommand{\figwidth}{0.8\linewidth}

\newcommand{\threefigwidth}{0.31\linewidth}
\newcommand{\alphahat}{\hat{\alpha}}

\usepackage{ulem}
\usepackage{xcolor}

\begin{document}

\title{Embedding method for the scattering phase in strongly correlated quantum dots}

\author{Rafael A. Molina}

\address{Instituto de Estructura de la Materia, CSIC, Serrano 123, 28006 Madrid, Spain.}

\ead{rafael.molina@csic.es}

\author{Peter Schmitteckert}

\address{Institute of Nanotechnology, Karlsruhe Institute of Technology, 76344 Eggenstein-Leopoldshafen, Germany}
\address{Center of Functional Nanostructures, Karlsruhe Institute of Technology, 76131 Karlsruhe, Germany}

\author{Dietmar Weinmann and Rodolfo A. Jalabert} 

\address{Institut de Physique et Chimie des Mat{\'e}riaux de
Strasbourg, UMR 7504, CNRS-UdS, \\
23 rue du Loess, BP 43, 67034 Strasbourg Cedex 2, France}



\author{Philippe Jacquod}

\address{Physics Department, University of Arizona\\
1118 E.\ Fourth Street, P.O. Box 210081, Tucson, AZ 85721, USA}
\address{D\'epartement de Physique Th\'eorique
Universit\'e de Gen\`eve
24, Quai Ernest Ansermet
1211 Gen\`eve, Switzerland}

\begin{abstract}
The embedding method for the calculation of the conductance through interacting systems connected 
to single channel leads is generalized to obtain the full complex transmission amplitude that 
completely characterizes the effective scattering matrix of the system at the Fermi energy.
We calculate the transmission amplitude  as a function of the gate potential 
for simple diamond-shaped lattice models of quantum dots with nearest neighbor interactions. 
In our simple models we do not generally observe an interaction dependent change in the number 
of zeroes or phase lapses that depend only on the symmetry properties of the underlying lattice. 
Strong correlations separate and reduce the widths of the resonant peaks while preserving the 
qualitative properites of the scattering phase. 

\end{abstract}

\section{Introduction}
The scattering theory of quantum transport has demonstrated to be a very successful tool for
the study and design of novel nanodevices. The information about the transport properties of 
a quantum mechanical system is encoded in the complex transmission amplitude. The conductance
of such a system in a two-terminal configuration can be obtained from the trace of the transmission
matrix $t$ multiplied by its Hermitian conjugate $t^{\dagger}$ according to the Landauer-B\"uttiker 
formula $G=(e^2/h) \mathrm{Tr} \left( t^{\phantom{\dagger}}t^{\dagger} \right)$ \cite{Landauer70,Buttiker86}. 
In a single channel device the conductance reduces to the modulus of the transmission probability $T$
through the channel and is easily measurable in experiments. On the other hand, even if the phases 
of the scattering matrix of a mesoscopic system show an interesting behaviour \cite{JP}, their 
direct measurement cannot be done. One possible way to extract the phase of 
the transmission coefficient is to embed the system to measure in one of the arms of an 
Aharonov-Bohm (AB) interferometer. Under the appropiate conditions it was shown to be possible to 
extract the transmission phase from the phase of the conductance oscillations of the whole 
structure as a function of the magnetic flux threading the AB ring \cite{Yacoby95,Schuster97}.

The pioneering AB experiments that tried to measure the transmission phase of transport through
a quantum dot in the Coulomb Blockade (CB) regime used a two-terminal geometry. In this situation, 
Onsager relations due to conservation of current and time-reversal symmetry imply that the 
conductance must be an even function of the magnetic flux and the measured transmission phase 
can only be $0$ or $\pi$ \cite{Buttiker86,Onsager31}.
Later experiments used a six-terminal configuration with additional leads that 
modify the reprocity relations for transmission coefficients \cite{Schuster97}. The obtained transmission phase presented a smooth 
increase of $\pi$ as a function of a gate potential in the dot $V_g$ when a resonance peak was 
crossed while a sudden jump of $-\pi$ (a phase lapse following the experimentalists convention) 
was observed between resonances.
In a more recent experiment the transition between this regime, termed universal regime, and a 
mesoscopic regime where phase lapses and resonances occur randomly with respect to each other 
was studied as a function of the number of electrons in the dot \cite{Avinum05}.

These observations have given rise to a whole body of literature devoted to study the transmission 
phase of quantum dots. Many theoretical works \cite{Baltin99,Hackenbroich01,Aharony02,Golosov06,Bertoni07} 
were concerned with explaining the experimental results, in particular, the universal regime for 
the transmission phase which is at odds with the standard models for CB phenomena. 
Recent progress in that direction was obtained studying the statistical properties of the parity 
correlations between succesive wave functions in ballistic chaotic quantum dots \cite{Molina11}. 
Another interesting path followed for the analysis of the experimental results is a many-body 
extension of the constant-interaction model (CIM) for the description of resonances in quantum 
dots \cite{Karrasch07,Karrasch07b,Molina11b}. In the regime where the distance between resonances 
is much smaller than the width of the resonant levels Karrasch {\em et al.} found that the number 
of phase lapses could be increased by moderate interactions. Although the results obtained do not 
seem to be generic, it has a lot of theoretical and practical interest to understand the effects
of strong correlations on the behavior of the full complex scattering quantities that characterize 
the electronic transport properties.

A complete theoretical understanding of transport through systems with strong correlations is generally missing, however. The problem can be formulated using the Keldysh approach \cite{Meir92,Datta92}. 
However, the calculations using Keldysh Green functions for interacting systems can usually only be 
performed within some approximation \cite{Schoeller00}.
There are also many numerical methods in the literature but none are without problems. The more 
generally applicable ones rely on extensive real time simulations, see \cite{Branschadel10} and 
references therein. 
However, the computing power needed for the application of these methods limits their practical 
implementations for strongly correlated systems. A more recent approach is
based on the exact diagonalization of teh full but isolated,
many-body system, and accordingly treats tunneling perturbatively
\cite{Bergfield09}. It obviously
is restricted to small systems.
Other types of approaches rely on the calculation of an equilibrium quantity that can be related 
to the conductance. Although these approaches are usually limited to the linear response regime 
they can be extremely useful to understand the effects of strong correlations on electronic 
transport \cite{embedding1,embedding2,embedding3,embedding4,embedding5,embedding6,othermethods}.
The purpose of this work is to present one of these methods, the embedding method, in a 
selfcontained manner. 
This numerical method in combination with Density Matrix Renormalization Group (DMRG) allows the 
computation of the transmission amplitude of a system with strong correlations attached to 
one-dimensional leads. As an illustration, we apply the method to simple lattice models of 
small quantum dots. 

\section{Embedding method for the transmission through a quantum dot}
\label{Sec:embedding}

During the last decade, the embedding method has been used quite successfully to calculate the 
transmission of electrons through simple lattice models of strongly correlated systems 
\cite{embedding1,embedding2,embedding3,embedding4,embedding5,embedding6}.
The system under consideration is embedded in a non-interacting ring. The
ground state properties of the combined system pierced by a magnetic flux 
can be related to the scattering properties of the embedded system.
In this section, we show the main steps of the derivation of the
method. We follow closely the approach that allows to extract the 
modulus of the transmission amplitude presented in Appendix A of Ref.\ \cite{molina04}. 
Moreover, we extend the method to calculate also the phase of the transmission amplitude. 
We will limit ourselves to the case of a single channel in the leads which is the simplest one
although generalizations of the method to more than one channel have been 
reported \cite{twochannels}.
The derivation is only rigorously valid for non-interacting impurities. However it was
shown numerically that an interacting system in the limit of infinite lead length can
be described by an effective scattering matrix whose matrix elements are accessible
via the embedding method \cite{molina04}.

\begin{figure}
\centerline{\includegraphics[width=0.5\linewidth]{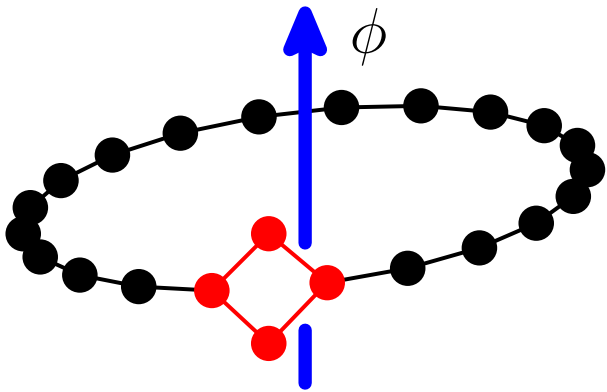}}
\caption{\label{fig:ringscheme}
Sketch of the system considered within the embedding approach: The interacting region of length 
$L_s$ (red) is embeded in a one-dimensional ring formed by a non-interacting lead (black) of 
length $L_L$. The ring is threaded by an Aharonov-Bohm flux $\phi$. The interacting system we 
consider consists of four sites in a diamond-shaped configuration with nearest-neighbor 
interaction.}
\end{figure}
We can write the quantization condition of the single-particle states of the ring in terms of 
the transfer matrix of the non-interacting part (the lead) and the transfer matrix of the 
scatterer (see the sketch in Fig. \ref{fig:ringscheme}) as
\begin{equation}
\mathrm{det}\left(I-M_\mathrm{L} M_\mathrm{S}\right)=0 \ ,
\label{quantcont}
\end{equation}
where $M_\mathrm{S}$ and $M_\mathrm{L}$ are the 
transfer matrices of the system and the lead, respectively. 
In the presence of time-reversal symmetry, the transfer matrix of a 
one-dimensional scatterer can be expressed in terms of three 
independent angles $\hat{\alpha}$, $\theta$ and $\varphi$:
\begin{equation}
\begin{aligned}
M_\mathrm{S} &= \left(\begin{array}{cc}
1/t^* & r^*/t^*\\
r/t & 1/t
\end{array}\right) \\ 
&= \frac{1}{\sin{\varphi}}
\left(\begin{array}{cc}
e^{i \hat{\alpha}}/\sin{\theta} & -i\cot{\theta}+\cos{\varphi} \\
i\cot{\theta}+\cos{\varphi} & e^{-i \hat{\alpha}}/\sin{\theta}
\end{array}\right)
\, , 
\end{aligned}
\label{eq:ms}
\end{equation}
where the two components correspond to right and left moving particles
while $r$ and $t$ are the reflection and transmission amplitudes,
respectively. The transmission amplitude is given by 
$t= e^{\mathrm{i} \hat{\alpha}} \sin{\theta} \sin{\varphi}=\left|t \right| \exp{\left( \mathrm{i} \alpha \right)}$.
This relation between the transmission phase $\alpha$ and the scattering phase $\hat{\alpha}$ was at
the origin of some misunderstandings about the interpretation of the experimentally observed
phase lapses until the issue was clarified by Taniguchi and B\"uttiker \cite{Taniguchi99}. 
$\alpha$ is the relevant experimental quantity and is not well defined when there is a zero of 
the transmission, resulting in a phase lapse whenever the transmission vanishes changing sign 
as a function of some parameter. $\hat{\alpha}$ is the relevant phase for the observance of the 
Friedel sum rule and only presents trivial $2\pi$ jumps. 

The transfer matrix of a lead of length $L_\mathrm{L}$ for a state with wave number $k\ge0$ reads
\begin{equation}
M_\mathrm{L} = \exp{(\mathrm{i} \Phi)} \left(\begin{array}{cc}
\exp(\mathrm{i}kL_\mathrm{L}) & 0\\
0 & \exp(-\mathrm{i}kL_\mathrm{L})
\end{array}\right),
\label{eq:ml}
\end{equation}
taking into account that the flux threading the ring can be included by twisting the
boundary condition of the system by a gauge transformation. Note that the scatterer embedded 
in the ring is not affected by this artificial flux such that we always work at zero magnetic 
field in the system of interest.

Inserting the transfer matrices (\ref{eq:ms}) and (\ref{eq:ml}), the eigenvalue
condition (\ref{quantcont}) yields
\begin{equation}
\cos(\Phi)=\frac{1}{\left| t(k) \right|}\cos\big( kL+\delta \hat{\alpha} (k) 
\big) \ .
\label{eq:quantcont2}
\end{equation}
The phase shift 
$\delta \hat{\alpha} = \hat{\alpha} - k L_\mathrm{S}$ is the phase of the scattering region relative
to a perfect lead of the same length $L_\mathrm{S}$. The solution of 
(\ref{eq:quantcont2}) yields the quantized momenta $k$ of the energy 
eigenstates in the lead.

Since both $t$ and $\delta \hat{\alpha}$ are functions of $k$, it is in 
general impossible to obtain an analytic solution of (\ref{eq:quantcont2}). 
However, one can study the asymptotic limit of large $L$, which was done by Gogolin and 
Prokof'ev in their study of the persistent current of a one-dimensional ring with a defect 
\cite{gogolin}. We use a general dispersion relation $\epsilon(k)$ in the lead that allows 
us to discuss continuum and tight-binding models at the same time.

The eigenvalue condition (\ref{eq:quantcont2}) can be rewritten as
\begin{equation}
k = k^0_n + \frac{1}{L} f_{\pm}(k,\Phi) \ .
\label{eq:eveq}
\end{equation}
Here, $k^0_n=2\pi n/L$ with $n\ge 0$ denotes the eigenvalues in the case of 
perfect transmission with $\vert t\vert=1$ and $\delta \hat{\alpha}=0$. Following the 
notation of Ref.\ \cite{gogolin}, we introduce the function
\begin{equation}
f_{\pm}(k,\Phi)= \pm \mathrm{Arccos} \left( \left|t(k) \right|
\cos \Phi \right) - \delta \hat{\alpha}(k) \  .
\end{equation}
Arccos denotes the principal branch of the inverse cosine function that takes values in 
the interval $[0,\pi]$. As $k$ should be positive, $f_-(k,\Phi)$ cannot be used for the 
case $n=0$. The splitting of the solutions of (\ref{eq:eveq}) corresponding to ``+'' and 
``-'' cannot exceed the spacing $2\pi/L$ between the $k_n^0$, provided that
$\delta\hat{\alpha}(k)$ is smooth on this scale. This is the case in the limit $L\to\infty$ 
and ensures that the order of the solutions with respect to energy is given by $n$. 

Iterating (\ref{eq:eveq}) and expanding $f_{\pm}$ for large systems, we obtain the expansion 
\begin{equation}\label{eq:kn}
\begin{aligned}
k_n^{\pm}= & k^0_n+\frac{1}{L}f_{\pm}(k^0_n,\Phi)\\
&+\frac{1}{L^2}f_{\pm}(k^0_n,\Phi)
\left( \frac {\partial f_{\pm}(k,\Phi)} {\partial k} \right)_{k=k_n^0} 
+  O \left(\frac{1}{L^3}\right)
\end{aligned}
\end{equation}
for the solutions of (\ref{eq:eveq}) in powers of $1/L$. 
Such an expansion is problematic in the vicinity of resonances where
there is a rapid variation of the function $f$ with $k$. In that case,
the expansion is valid only for sufficiently large $L$.

We now calculate the ground state energy of the system as a function of the 
flux to order $1/L^2$. Using (\ref{eq:kn}), the expansion of the 
one-particle energies in powers of $1/L$ can be shown to be
\begin{equation}
\begin{aligned}
\epsilon(k^\pm_n) =& \epsilon(k^0_n)+\frac{1}{L}
\left(\frac{\partial \epsilon}{\partial k} f_{\pm}(k,\Phi)\right)_{k=k_n^0}\\
&+\frac{1}{2L^2}\frac {\partial } {\partial k} 
\left(\frac{\partial \epsilon}{\partial k} f_{\pm}^2(k,\Phi)\right)_{k=k_n^0}
+ O\left(\frac{1}{L^3}\right)\, .
\label{eq:epsilon2}
\end{aligned}
\end{equation}

We consider the simplest case of an odd number of particles in the ring. The
leading order results for even number of particles give rise to the same final 
expression for the calculation of the transmission modulus \cite{molina04}.
For an odd number of particles $N$ in the ring, all occupied states $n$ come in 
pairs ([$n$,-] and [$n$,+]), except for the one corresponding to $n=0$. 
The total ground state energy is
\begin{equation}
\begin{aligned}
&E_0^\mathrm{odd}(\Phi)= \epsilon(k_0^+) 
+ \sum_{n=1}^{n_\mathrm{F}} [\epsilon(k_n^+) + \epsilon(k_n^-)] \\
&\;= \epsilon(0)
+\frac{1}{2L^2}\left(\frac{\partial^2\epsilon}{\partial k^2}
\left[\mathrm{Arccos}(|t|\cos\Phi)-\delta \hat{\alpha}\right]^2\right)_{k=0}\\
&\quad + \sum_{n=1}^{n_\mathrm{F}} \left\{ 2 \epsilon(k_n^0) - 
\frac{2}{L} \left(\frac{\partial \epsilon}{\partial k}\right)_{k=k_n^0} 
\delta \hat{\alpha} (k_n^0) \right. \\ 
& \left. \quad+ \frac{1}{L^2}
\frac{\partial}{\partial k} \left(\frac{\partial\epsilon}{\partial k}
\left[\mathrm{Arccos}^2(|t|\cos\Phi)+\delta\hat{\alpha}^2\right]\right)_{k=k_n^0}
\right\} + O \left(\frac{1}{L^3}\right)\,  .  
\label{genergy}
\end{aligned}
\end{equation}
The sum runs up to $n_\mathrm{F}=(N-1)/2$.
We have assumed $(\partial \epsilon/\partial k)_{k=0}=0$ and kept all terms 
which can give rise to contributions up to order $1/L$. 
The first term in the sum is the ground state energy in the absence of scattering. 
It is proportional to $L$ when $N$ is of order $L$. The second term represents
the energy change due to the scattering potential and is of order 1. 
From this second term, we can obtain the scattering phase shift by changing the number of 
particles as it is discussed below. The third term is the leading flux-dependent 
correction which allows to obtain the modulus of the transmission.  

We start by considering the leading order terms for the calculation
of the modulus of the transmission (see App.\ A of Ref.\ \cite{molina04} for
next order corrections and for the case of an even number of particles).
Converting the sums over $n$ into integrals, these flux-dependent 
contributions can be expressed as 
\begin{equation}
\label{eq:int1}
\begin{split}
&\frac{1}{2 \pi L} \int\limits_{\pi/L}^{k_F+\pi/L}\mathrm{d}k
\frac{\partial}{\partial k} \left(\frac{\partial\epsilon}{\partial k}
\mathrm{Arccos}^2(|t|\cos\Phi)\right)\\
&\qquad= \frac {\hbar v_\mathrm{F}} {2 \pi L}\mathrm{Arccos}^2\big(|t(k_F)|
\cos(\Phi)\big) 
+ O\left(\frac{1}{L^2}\right). 
\end{split}
\end{equation}
Here, $k_\mathrm{F}=2\pi n_\mathrm{F}/L$ is the Fermi wave 
number and $v_\mathrm{F}=(\partial\epsilon/\hbar\partial k)_{k=k_\mathrm{F}}$ 
is the Fermi velocity. The leading flux-dependent 
term of the ground state energy is 
\begin{equation}
E_0^{\mathrm{odd}(1)}(\Phi)= 
\frac {\hbar v_\mathrm{F}} {2 \pi L}
\mathrm{Arccos}^2\big(|t(k_F)|\cos(\Phi)\big),
\end{equation}
where the superscript indicates the order of the result in $1/L$.
In order to calculate the transmission modulus we need  
the leading order of the charge stiffness
\begin{equation}\label{eq:stiffleadingapp}
\begin{aligned}
D^{(1)}&=
-\frac{L}{2}\left(E_0^{\mathrm{odd}(1)}(0)-E_0^{\mathrm{odd}(1)}(\pi)\right)\\
&= \frac{\hbar v_\mathrm{F}}{2} 
\left[\frac{\pi}{2}-\mathrm{Arccos}(|t(k_F)|)\right]\, .
\end{aligned}
\end{equation}   
This last result is independent of the parity of the number of particles. 
Then, the transmission modulus can be obtained as
\begin{equation}\label{eq:modt}
\left|t(E_F)\right|^2=\lim_{L\to\infty} \sin^2 \left(\frac{\pi D}{2 D^0}\right),
\end{equation}
where $D^{0}$ is the charge stiffness of a clean ring without scatterer.

We now keep the flux fixed. Then, the lowest order change of the ground state 
energy (9) in $1/L$ with repect to the change of the particle number $N$ 
depends on 
$\delta\hat{\alpha}$. It is thus possible to express the scattering phase as
\begin{equation}\label{eq:phase-general}
\delta\alphahat(k_\mathrm{F})=-\lim_{L\to\infty} \frac{L}{2}
\left(\frac{E(N)-E(N-2)-2\epsilon(k_{n_\mathrm{F}}^{(0)})}
{\left.\mathrm{d}\epsilon/\mathrm{d}k\right|_{k=k_{n_\mathrm{F}}^{(0)}}}\right)
\end{equation}
in terms of the many-body energies at particle numbers $N$ and $N-2$ and the 
properties of the dispersion relation at the Fermi energy. In the special case of a 
tight binding chain with unit hopping and lattice spacing at half filling, this formula 
reduces to
\begin{equation}\label{eq:phase}
\delta\alphahat(k_\mathrm{F})= - \lim_{L\to\infty} \frac{L}{4}
\left[E(L/2+1)-E(L/2-1)\right]\, .
\end{equation}
Eqs.\ \ref{eq:phase-general} 
and \ref{eq:phase} show that the scattering phase is related to the change 
in total ground state energy in large rings when the particle number is 
changed. It is therefore possible to use the embedding method to 
extract the scattering phase.

From the embedding method, we can thus obtain $\alphahat$. However, the
quantity of interest from the experimental point of view is $\alpha$ with
its phase lapses coming from the different branches of the relation
$|t|\exp(i\alpha)=\sin\theta\sin\varphi\exp(i\hat{\alpha})$.
To achieve this goal, we need to access the sign of $\varphi$. 
For that purpose we go back to the general expression
\eqref{eq:ms} of the transfer matrix and see that in order to have 
left-right symmetry we need $r=r'$, and therefore 
$\sin{\theta}\cos{\varphi}=0$.

We have three possibilities: i) $\sin{\theta}=0$, ii) $\varphi=\pi/2$, 
iii) $\varphi=-\pi/2$. Thus, in a left-right symmetric case we have 
two branches ($\sin{\varphi}=1$ and $\sin{\varphi}=-1$) and we can pass 
from one to the other only when $t=0$. 

The embedding method allows to follow the branch-switching since the
phase sensitivity is given by \cite{molina04}
\begin{equation}\label{eq:phsen}
\Delta E =\big( E(\pi)-E(0)\big)=\frac{\hbar v_\mathrm{F}}{L}
\left[\frac{\pi}{2}-\mathrm{Arccos}(\sin{\theta}\sin{\varphi})\right] \, ,
\end{equation}
and therefore
\begin{equation}\label{eq:sinthsinphi}
\sin{\theta}\sin{\varphi}=
\sin{\left(\Delta E \frac{L}{\hbar v_\mathrm{F}}\right)}=
\sin{\left(\frac{\pi}{2} \frac{\Delta E}{|\Delta E^{(0)}|}\right)} \, ,
\end{equation}
where $\Delta E^{(0)}$ is the phase sensitivity of a perfectly
transmitting scatterer. Therefore, the sign changes of $\Delta E$ determine
the branch-switch of $\varphi$ and thus the jumps in $\alpha$. One should
also notice that there can also be transmission zeroes without phase lapses
in cases when there is a zero of $\Delta E$ without sign change. 
This possibility does occur for particular values of
the parameters in some of the models we have
analyzed.


For a strictly one-dimensional
system, the sign of $\Delta E$ is fixed by Leggett's theorem 
\cite{leggett,waintal08}. 
For an odd number of particles we are in the branch of positive $\varphi$, $t$
never vanishes, and there cannot be branch-switches. For a quasi-one
dimensional scatterer this is no longer true, there can be
parameter values where $t$ vanishes and branch-switches appear. 
The embedding method for the phase of the transmission has been checked for
a one-dimensional chain and its predictions compared with the Friedel sum rule.
Its validity has been confirmed for interacting scatterers attached to reservoirs via
non-interacting leads, where it describes
the effective one-body scattering at the Fermi-energy \cite{Molina11b,molina04}.



\section{Numerical results for diamond-shaped quantum dots}
\label{Sec:diamond}

A one-dimensional scatterer does not exhibit zeroes of the transmission and 
thus no phase lapses are observed. We therefore study a minimal model 
of a scatterer in which transmission zeroes occur, and where phase lapses are 
expected. In this context, the diamond lattice quantum dots were introduced
by Levy-Yeyati and B\"uttiker as a simple example to understand the physics behind the
phase lapses of the transmission \cite{levy00}. In the non-interacting case the form of the
transmission as a function of the gate voltage $V_g$ can be obtained analytically \cite{levy00}.
The results for the non-interacting case can serve as a guide for the convergence properties
of the embedding method for quantum dot lattice models beyond one dimension.

The Hamiltonian of the system under study including interactions reads
\begin{eqnarray}
H &=& -t \left( c_{1}^{\dagger}c_{2}^{\phantom{\dagger}} +  c_{1}^{\dagger}c_{3}^{\phantom{\dagger}} +  c_{2}^{\dagger}c_{4}^{\phantom{\dagger}} +  c_{3}^{\dagger}c_{4}^{\phantom{\dagger}} + \mathrm{H.C.} \right) + \epsilon n_2 - \epsilon n_3 \nonumber \\ 
&& + V_g \sum_{i=1}^4 n_i + U \left( \tilde{n}_1\tilde{n}_2 + \tilde{n}_1 \tilde{n}_3 + \tilde{n}_2 \tilde{n}_4 + \tilde{n}_3 \tilde{n}_4 \right), 
\label{eq:Hamiltonian}
\end{eqnarray}
where $\tilde{n}_i=n_i-1/2$. The nearest neighbor interaction is added in that form in order to 
enforce charge neutrality at half-filling. A sketch of the lattice used in the model is shown in 
Fig.\ \ref{fig:ringscheme}. 
Site $1$ is connected to the left lead and to sites $2$ and $3$ while site $4$ is connected to the 
right lead and also to sites $2$ and $3$. The single particle energies of sites $2$ and $3$ are the 
same but with opposite signs while the single particle energies at the sites connected to the leads 
$1$ and $4$ are taken to be $0$. This symmetry ensures that a zero of the transmission appears 
exactly at $V_g=-2U$.

\begin{figure}
\centerline{\includegraphics[width=\figwidth]{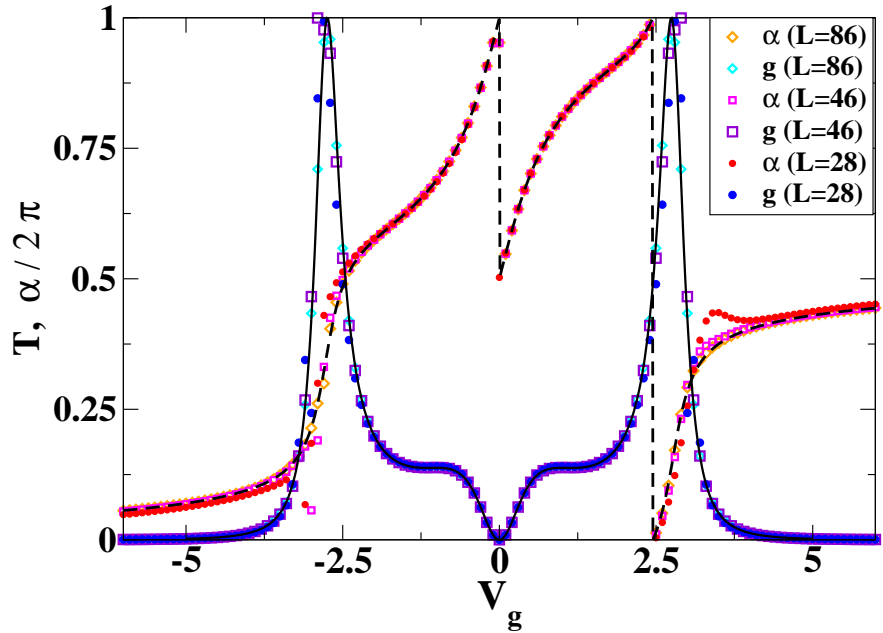}}
\caption{\label{fig:embni}Transmission probability and transmission phase for the non-interacting diamond dot model 
with $\epsilon=2$ calculated using the embedding method. The solid lines represent the exact results.}
\end{figure}

In the non-interacting case with $\epsilon=2$ the conductance for this model has two symmetric 
resonant peaks  at $V_g=2.75$ and $V_g=-2.75$. The zero of the transmission is 
associated with a phase lapse at $V_g=0$. There are two more symmetric resonances around the 
transmission zero at $V_g=-1.01$ and $V_g=1.01$ but corresponding to very badly coupled wave-functions 
so they hardly surpass the transmission background from the bigger resonances. We compare the exact 
results to calculations using the embedding method. For obtaining results from the embedding method an 
extrapolation to infinite lead lengths is needed. 
A finite size scaling analysis of the results is made using increasing lead lengths and linear 
or quadratic extrapolation formulas as explained in detail in Ref.\ \cite{molina04}. This procedure 
is performed for the transmission modulus and for the transmission phase in exactly the same fashion. 
We show in Fig.\ \ref{fig:embni} the values for $g$ and $\alpha$ obtained with total 
lengths of the system plus the auxiliary ring of $L=28,46,86$ and compare them with the exact result. 
The actual value of the length that we need to use in the extrapolation formulas (Eq.\ 
\ref{eq:stiffleadingapp} and Eq. \ref{eq:phase}) is $L-1$ due to the geometrical form of the system 
(sites 2 and 3 are just to alternative ways for the electron to go through the system and count as a 
single site for the purpose of the extrapolation formulas). The results for all the three cases
are very good around the zero of the transmission. There are small deviations close to the resonances 
for $L=28$ and $L=46$ while the results for $L=86$ coincide almost exactly with 
the exact results. The qualitative behavior is correctly reproduced even for the smallest ring size 
$L=28$. The deviation from the exact results close to the resonances is characteristic of the 
method when full convergence in the length of the auxiliary ring has not been reached. Similar 
deviations are observed when interactions are switched on. 

When $U>0$, the ground state properties of the ring including the diamond-shaped quantum dot 
are calculated using the Density Matrix Renormalization Group (DMRG) algorithm \cite{DMRG}. We keep 
up to 650 states in order to get a good accuracy in the calculation of the charge stiffness of the 
ring for the larger sizes. 

From very general arguments based on the Green's function approach to non-interacting transport it can 
be shown that a zero of the transmission appears between two consecutive resonant peaks depending on the
relative symmetry properties of the wave functions corresponding to the resonances with respect to the 
positions of the leads \cite{levy00}. If consecutive resonances have the same symmetry a zero appears 
between peaks and when they have opposite symmetry the zero is missing. The four resonances of the 
diamond quantum dot Hamiltonian Eq. \ref{eq:Hamiltonian} have even, odd, odd and even symmetry with 
respect to sites $1$ and $4$ so there is only one zero in the middle.

\begin{figure}
\includegraphics[width=\threefigwidth]{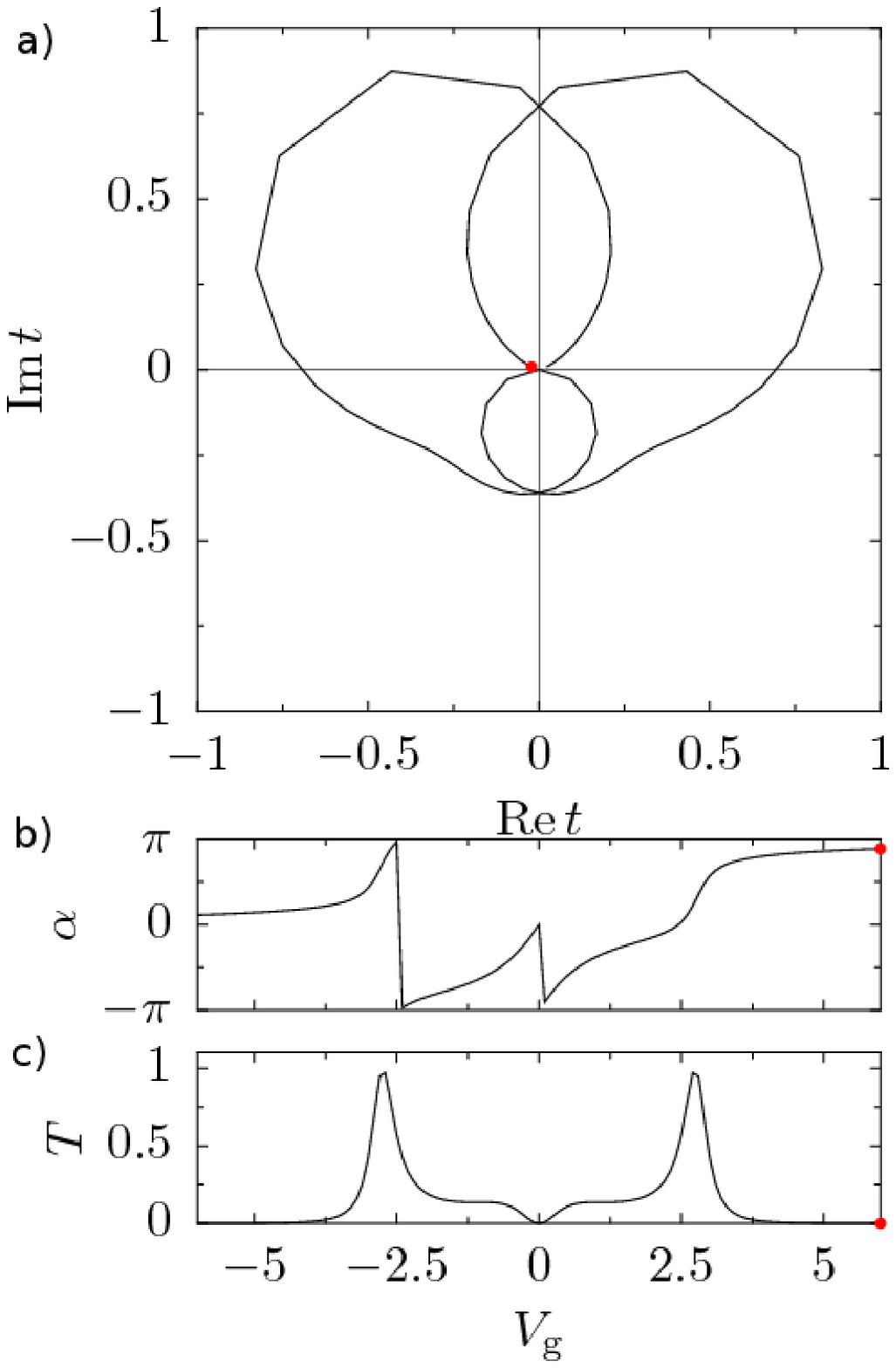}
\includegraphics[width=\threefigwidth]{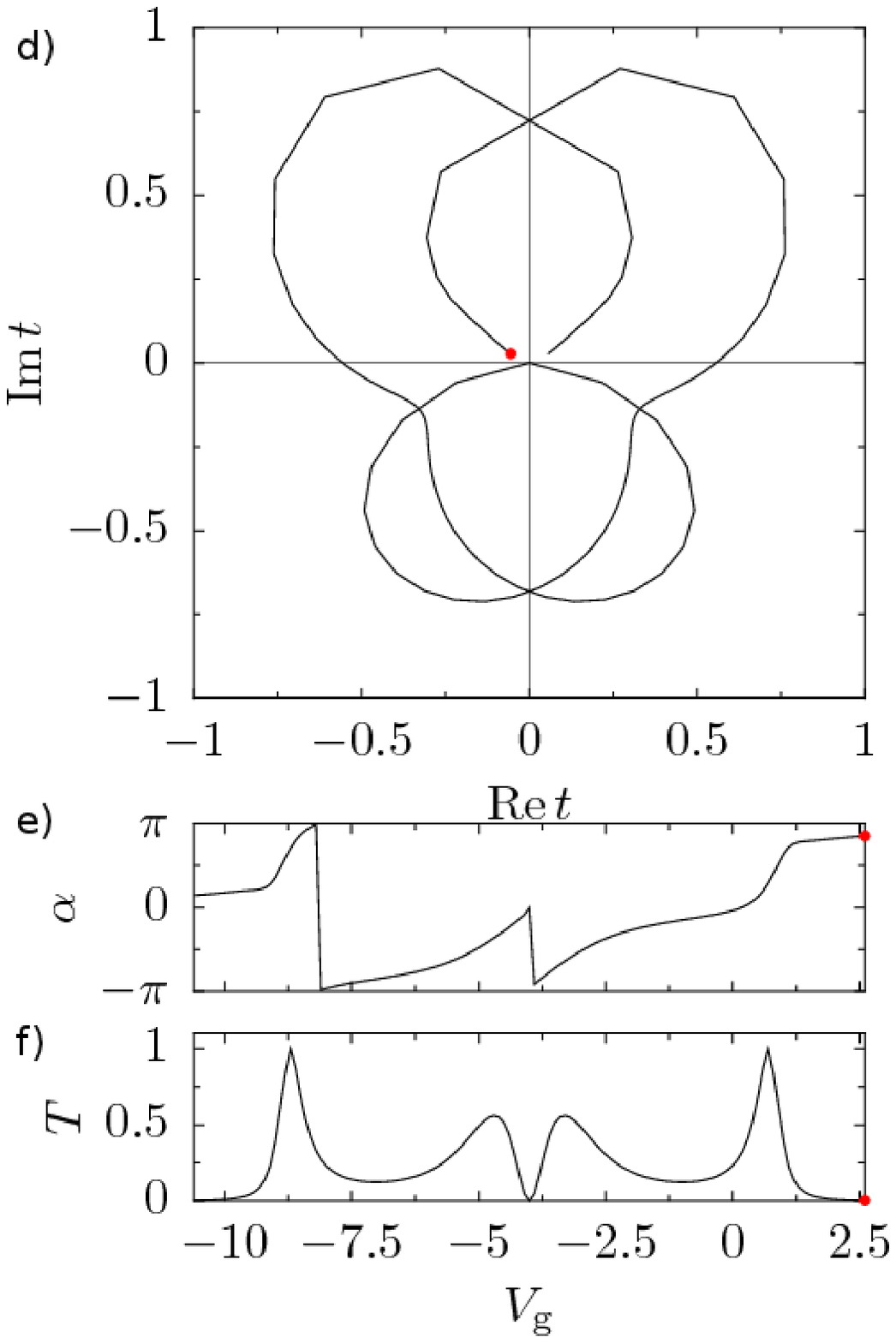}
\includegraphics[width=\threefigwidth]{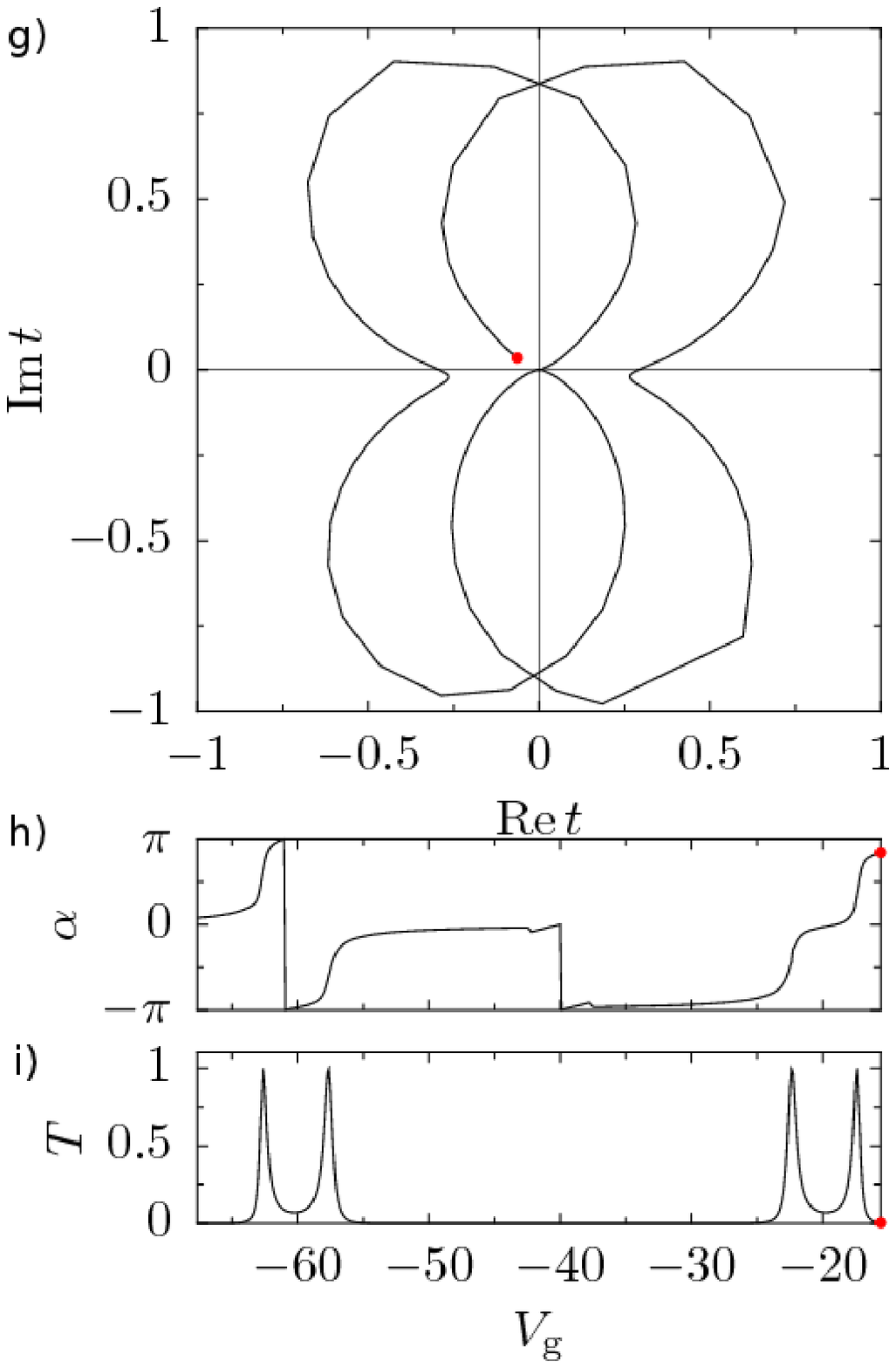}
\caption{\label{fig:rombou}
Results for the diamond quantum dot model with $\epsilon=2$ and different values of the interaction
strength $U$. The left, middle, and right panels show the
trajectory of the complex transmission amplitude (a, d, and g), transmission phase as a function of $V_g$ 
(b, e, and h)
and transmission probability $T$ as a function of $V_g$ (c, f, and i) for $U=0$, $U=2$, and $U=20$, respectively.
The red dot marks the same $V_g$ point in each of the three graphs with the same value $U$ so it is easier 
to follow the trajectory of the 
transmission amplitude in the complex plane.}
\end{figure}

The effect of the interactions is to change the position of the zero of the conductance (due to the 
charge neutrality requirements) and the position of the resonance peaks while deforming their overall 
shape. The resonances get more isolated and symmetric. However, the phase lapse of $\alpha$ when we go 
through the transmission zero is the same as in the non-interacting case. Within our model, The 
symmetries of the resonances are not modified by the many-body correlations that are induced
through the increase of the interaction strength $U$. In the different panels of Fig.\ \ref{fig:rombou} 
we show the behavior of the complex transmission coefficient as a function of $V_g$. We compare the 
exact results for $U=0$ with embedding method results calculated with an extrapolation up to $L=76$ for 
the cases of $U=2$ and $U=20$. 

Due to the symmetry of the problem, the transmission zero is always placed between the two groups 
of resonances. As we increase the interaction the symmetric resonances close to the zero of 
transmission that are barely visible in the case $U=0$ fully develop into lorentzian peaks and a 
wide gap opens between them. This is due to the transition between extended wave functions for 
small $U$ to spatially localized wave-functions for large $U$. However, there is no qualitative 
change in the behavior of the transmission phase as we increase the interaction $U$. The 
trajectories of the complex transmission amplitude as we vary $V_g$ shown in the top panel of the 
figure show similar features. 
The mechanisms proposed for the change of the number of phase lapses in interacting models like
population switching\cite{Karrasch07,Karrasch07b,Golosov07} are not at work in this simple geometrical 
model of a quantum dot precisely because the geometrical properties of the lattice dominate
the symmetry properties of the wave functions even in the many-body case.




\section{Conclusions}
\label{Sec:conclusions}

The embedding method for the calculation of the transmission through a strongly correlated system 
connected to non-interacting leads can be successfully extended to calculate also the phase of the 
transmission amplitude. The phase of the effective transmission amplitude of a correlated system 
connected to non-interacting one-dimensional leads can be extracted from differences of the ground 
state energy of an auxiliary ring at different electron numbers. 
We have illustrated the application of this method for the calculation of the full complex 
transmission coefficient of a diamond-shaped quantum dot with nearest-neighbor interactions. 
The calculations of the ground state energy differences have been performed using the DMRG 
algorithm that allows for great accuracy in the models considered. For the small systems that we 
treated in the work, interactions modify substantially the shape of the resonances although the 
number and position of the zeros between resonances are not changed and depend only on the symmetry
properties of the quantum dot model.Larger structures which represent a harder computational challenge
may present richer phenomena concerning the scattering phase. These and further issues will be explored
in future work.

\ack

We acknowledge support from the
Spanish MICINN through project FIS2009-07277, the NSF under
grant No DMR-0706319, the ANR through grant ANR-08-BLAN-0030-02,
and the Swiss NCCR MANEP.

\end{document}